# Long-range Ordering of Vibrated Polar Disks

C. A. Weber,[1] T. Hanke,[1] J. Deseigne,[2] S. Léonard,[3] O. Dauchot,[4,3] E. Frey,[1] and H. Chaté[3,5]

[1]*Arnold Sommerfeld Center for Theoretical Physics and Center for NanoScience, Department of Physics, Ludwig-Maximilians-Universität München, 80333 Munich, Germany*
[2]*Laboratoire de Physique, ENS de Lyon UMR 5672, 69007 Lyon, France*
[3]*Service de Physique de l'Etat Condensé, CEA-Saclay, URA 2464 CNRS, 91191 Gif-sur-Yvette, France*
[4]*EC2M-Gulliver, ESPCI-ParisTech and CNRS UMR 7083, 75005 Paris, France*
[5]*Max Planck Institute for the Physics of Complex Systems, Nöthnitzer Str. 38, 01187 Dresden, Germany*

Vibrated polar disks have been used experimentally to investigate collective motion of driven particles, where fully-ordered asymptotic regimes could not be reached. Here we present a model reproducing quantitatively the single, binary and collective properties of this granular system. Using system sizes not accessible in the laboratory, we show *in silico* that true long-range order is possible in the experimental system. Exploring the model's parameter space, we find a phase diagram qualitatively different from that of dilute or point-like particle systems.

Collective motion in driven or self-propelled particle systems is a topic of recent interdisciplinary interest [3–6]. Within physics, following the works of Vicsek *et al.* [7, 8] and Toner and Tu [9–11], most progress was achieved by studying microscopic models [7, 8, 12–24] and their continuous descriptions [9–11, 25–38]. For the simplest situation in which the surrounding fluid can be neglected ("dry flocking") and the sole interaction is some local effective alignment, a picture of basic universality classes has emerged, which connects models similar to the Vicsek model [7] to continuous theories of the Toner-Tu type [9–11, 25–31]. Among the landmark results are the possibility of true long-range orientational order in two dimensions, the generic presence of strong, long-range correlations [9–11, 25] and/or spontaneously segregated dense and highly ordered nonlinear structures in moving, ordered, fluctuating phases [29, 31].

These numerical and theoretical results still largely lack experimental confirmation. This is mostly due to the fact that decisive experimental tests must be performed on large numbers of objects under controlled conditions. The advent of experiments using purified proteins (motors, filaments, etc.) offers a promising playground [39–43], but another line of attack, for dry flocking, is to build on the experience of the granular physics community, and to shake man-made objects [1, 2, 44–48]. Recently, some of us have designed and studied the collective motion of vibrated polar disks, i.e. millimeter-size objects with a built-in oriented axis and a circular top metallic part rendering the particles isotropic with respect to collisions (Fig. 1a,b; [1, 2]). Large-scale collective streams and anomalous, "giant" number fluctuations were reported in collections of approximately a thousand disks moving on a carefully vibrated plate. Unfortunately, in this experiment — as in others involving man-made objects [44–48] — the number of particles used was still too small to reach asymptotic results. Moreover, the most ordered regimes that could be explored were probably close to the onset of collective motion, making it impossible to disentangle the properties of the ordered moving phase from those of the order-disorder transition.

In this work, we bypass the inherent difficulties of the experimental setup for vibrated polar disks by studying the system *in silico*: We construct a model for the motion and collisions of the polar disks of [1, 2], which accounts quantitatively for most of the known experimental properties at the single and pair interaction level. Our model also agrees well with observations at the collective level. This allows us to study system sizes and boundary conditions unreachable in the laboratory. We show that even in the most ordered regimes observed experimentally, no long-range collective motion exists. However, changing parameters only slightly we find ordered regimes which could be observed in the laboratory. Exploring the model's parameter space systematically, we discover a phase diagram qualitatively different from that of dilute or point-like particle systems [20, 29]. In particular, we find, at rather large packing fractions, "inverse bands" and a possibly direct transition from disorder to a "Toner-Tu" [9–11, 25] collectively moving phase.

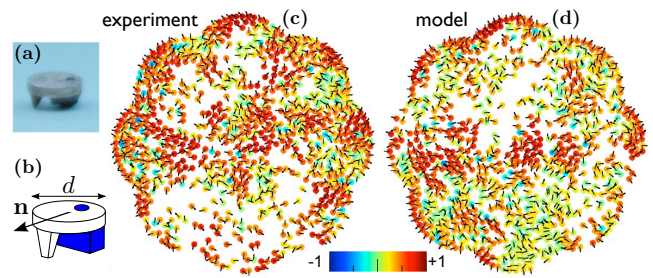

FIG. 1. (color online) **(a)** Photograph and **(b)** sketch of one polar disk, with particle's polarity **n** indicated. Typical snapshots in the petal-shaped geometry, for the experiment **(c)**, and the model **(d)**. Parallel [anti-parallel] particles are shown in red (+1) [blue (-1)]. For details refer to [49].

The polar disks (Fig. 1a,b) are vibrated between two plates. Rather than modeling their full three-dimensional dynamics, we describe their effective two-dimensional motion. Dictated by the experimental system the main new features of the model are: (i) the dynamics of the particle's intrinsic polarity with respect to their velocity is explicitly described, and (ii) no explicit alignment rules are employed, but collisions are modeled explicitly. Building on experimental observations, notably that single particles move backward for significant time periods with their velocity essentially antiparallel to their director, we were led to the following model: Particle $i$ is subject to a noisy acceleration along its *polarity* axis $\mathbf{n}^i$ (with anisotropic, intrinsic, "active" noise, respecting the particle's polar symmetry), balanced by an effective linear friction term along its *velocity* $\mathbf{v}^i = \frac{d}{dt}\mathbf{r}^i$, with $\mathbf{r}^i$ denoting the particle's coordinates. Particles $i$ and $j$ with $|\mathbf{r}^i - \mathbf{r}^j| < d$, where $d$ is the particle diameter, interact by means of a pairwise, inelastic, repulsive interaction force $\mathbf{F}_\epsilon^{ij}$, yielding the equations:

$$\frac{d}{dt}\mathbf{v}^i = [\mu + \eta_\parallel]\mathbf{n}^i + \eta_\perp \mathbf{n}_\perp^i - \beta \mathbf{v}^i + \sum_j \mathbf{F}_\epsilon^{ij}, \quad (1)$$

where $\mu$ and $\beta$ are constants giving rise to a stationary speed $v = \mu/\beta$, $\mathbf{n}_\perp^i$ is a unit vector perpendicular to $\mathbf{n}^i$, $\eta_{\parallel,\perp}$ represent Gaussian distributed white noises with zero mean, i.e. $\langle \eta_{\parallel,\perp}(t)\eta_{\parallel,\perp}(t')\rangle = 2D_{\parallel,\perp}\delta(t-t')$, where $D_{\parallel,\perp}$ denotes the corresponding diffusion constant. The interaction force $\mathbf{F}_\epsilon^{ij}$ is given by the established spring dash-pot model [50, 51], which, for hard particles, depends only a single parameter, the restitution coefficient $\epsilon$ [52].

Eq. (1) must be complemented by one governing the polarity of particles, which was observed to remain antialigned to the velocity during episodes of backward motion. In other words, when $\alpha^i = \angle(\mathbf{v}^i, \mathbf{n}^i)$, the angle between velocity and polarity, is acute, frictional interactions with the vibrating plate are assumed to rotate $\mathbf{n}^i$ towards $\mathbf{v}^i$, while for $|\alpha^i| > \pi/2$, $\mathbf{n}^i$ rotates towards $-\mathbf{v}^i$. We thus propose the following equation for the polarity angle $\phi^i$ [with $\mathbf{n}^i = (\cos\phi^i, \sin\phi^i)$]:

$$\frac{d}{dt}\phi^i = \zeta \sin\alpha^i \,\mathrm{sign}(\cos\alpha^i), \quad (2)$$

where $\zeta$ characterizes the strength of the coupling between polarity and velocity. This parameter is expected to be rather small given the observed persistence of $\mathbf{n}$ even when $\mathbf{v}$ changes sign abruptly.

To make contact with the experimental results, we rescale time $t \to t/\tau_0$, with $\tau_0$ the inverse of the vibration frequency $f = 115$ Hz [1, 2]. Length is measured in particle diameters $d$: $\mathbf{x} \to \mathbf{x}/d$. Our model possesses six parameters: $\mu$, $\beta$, $\zeta$, $D_\parallel$, $D_\perp$, and $\epsilon$. At fixed experimental vibration amplitude $\Gamma$, one parameter can be eliminated by matching the typical experimental speed with the model's velocity $v = \mu/\beta$. In the following, we use the experimental data gathered at the vibration amplitude $\Gamma = 2.7$, where the most ordered regimes have been observed, and for which $v = 0.025$ [1, 2].

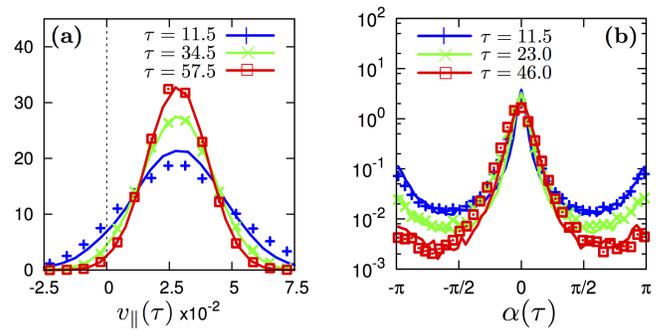

FIG. 2. (color online) **(a)** PDF of $v_\parallel$ and **(b)** PDF of the angle $\alpha = \angle(\mathbf{n}(t), \Delta\mathbf{r}(t+\tau))$ [lin-log] for selected values of the time increment $\tau$. Experimental data are indicated with symbols; model data are illustrated with lines.

We first analyze the single-particle dynamics in order to test the overall quality of the model and to estimate the remaining four parameters (i.e. $\beta$, $\zeta$, $D_\parallel$, $D_\perp$; the restitution coefficient $\epsilon$ only affects particle interactions). To find the best-matching set of parameters, we consider the following two quantities: the angular diffusion constant $D_\phi$ and the ratio of the displacement fluctuations parallel and perpendicular to the polarity (definitions see [53]). Scanning the four dimensional parameter space, we select a best-matching parameter set for which both quantities agree with the experimental value within an accuracy of ±30%. This is approximately equal to the imprecision arising due to different preparations of the experimental setup (see [49] for more information). In spite of this modest accuracy, the model captures quantitatively the observed experimental particle dynamics: We compare the distributions of the parallel displacements normalized by $\tau$, denoted as $v_\parallel(\tau) = \Delta r_\parallel/\tau$ ($\Delta r_\parallel$ is defined in [53]), and of the angle $\alpha(\tau) = \angle(\mathbf{n}(t), \mathbf{r}(t+\tau) - \mathbf{r}(t))$ to those recorded experimentally. We find a very good agreement for all values of $\tau$ considered (Fig. 2). Note that, as expected, the particles exhibit backward motion for significant time periods (tails in the negative sector in Fig. 2a, and peaks at $\pm\pi$ in Fig. 2b).

We now turn to binary collisions, for which the restitution coefficient $\epsilon$ must be chosen. The following results are presented for $\epsilon = 0.4$, but we observed that changing $\epsilon$ in the range ±30% does not influence collision properties significantly [49]. Experiments have revealed that one "encounter" typically involves many successive collisions, where the particles bounce back without turning their polarity much, so that they quickly collide again. These encounters last for a finite time and take place over some finite spatial extension. It was found experimentally that they are well delimited using the following criterion: an encounter starts when two particles get closer than some threshold collision distance, i.e. $|\mathbf{r}^i - \mathbf{r}^j| \le d_c = 1.7$, and their polarities point "inwards", i.e. $|(\mathbf{r}^i + \mathbf{n}^i) - (\mathbf{r}^j + \mathbf{n}^j)| \le |\mathbf{r}^i - \mathbf{r}^j|$ [2]. An encounter ends either when particles are separated by more than $d_c$, or their polarities point "outwards". We have used the same criterion for our model. Fig. 3 depicts the results of a scattering study for the experimental setup and the model. Thousands of binary encounters (hereafter called collisions for simplicity) were recorded, and the outgoing relative angle $\theta_{\rm out}$ of the two particles plotted against their incoming relative angle $\theta_{\rm in}$, the impact parameter $b \in [0, 1]$ [51] is shown as color code (Fig. 3a,b). The model data shows a striking agreement with the results measured in the experiments: most collisions actually leave the polarities unchanged ($\theta_{\rm out} \simeq -\theta_{\rm in}$), and a minority of them align the particles almost perfectly ($\theta_{\rm out} \simeq 0$). We estimated the fraction of polar aligned events [54], finding 0.14 for the model and 0.18 for the experiment. The model also matches the distribution of head-on ($b \approx 0$) and glancing ($b \approx 1$) collision events. We further determined the PDF of the duration of collisions $\tau_{\rm col}$ as well as that of their spatial extension $\ell_{\rm col}$, given by the center of mass displacement. The model reproduces the observed exponential distribution of $\tau_{\rm col}$ quantitatively, while it fails to reproduce the roughly algebraic decay of $\ell_{\rm col}$ (but nevertheless gives a correct mean extension)[55].

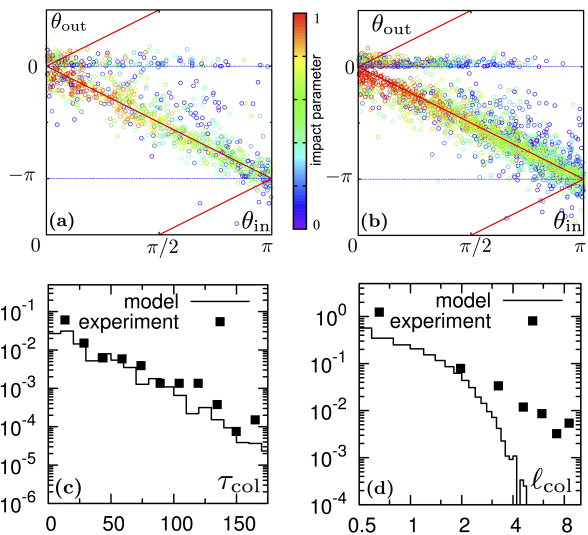

FIG. 3. (color online) Scatter graph $\theta_{\rm in} - \theta_{\rm out}$ for the experiment (**a**), and our model (**b**). Values of the impact parameter $b$ are indicated by the color bar. PDF of the duration $\tau_{\rm col}$ (**c**) [lin-log], and the extension $\ell_{\rm col}$ of a collision (**d**) [log-log].

We performed simulations using the same flower-shaped geometry (Fig. 1), and number of particles ($N = 890$) as in the experiment [1, 2]. For the parameter values matching the single particle dynamics and binary collisions (for vibration amplitude $\Gamma = 2.7$), we observe, as in the experiments, fairly large, polar aligned, moving clusters (Fig. 1c,d, for videos refer to [49]). However, the order parameter $\psi(t) = \frac{1}{M(t)} |\sum_{i \in {\rm ROI}} \mathbf{n}^i|$, with $M(t)$ denoting the number of particles currently located within the central "region of interest" (ROI) of radius 10, is typically smaller than in the experiment (Fig. 4a). The effective packing fraction observed in the ROI is found to be very close to that of the experiment ($\phi \simeq 0.39$, whereas the nominal packing fraction is 0.47), indicating that particles accumulate at the boundary in the model as well. Running the model at $\phi = 0.39$ in a box of approximately the same size but with periodic boundary conditions —a privilege of the *in silico* approach— yields only a marginally larger average polarization (Fig. 4a): a frustration-free geometry is unable to restore enough order.

We also ran the model in square periodic domains of linear size $L$ at the nominal packing fraction $\phi = 0.47$, and then found order being slightly stronger than in the experiment (Fig. 4a). Nevertheless, increasing system size $L$, we observe that the overall order parameter $\langle \Psi \rangle_t = \langle \frac{1}{N} | \sum_{i=1}^{N} \mathbf{n}^i | \rangle_t$ decreases first rather slowly, then faster (Fig. 4b, inset). Thus, no true long-range order is present at the exact conditions probed experimentally. In fact, the correlation length can be estimated by the kink in the average polarization as a function of system size (inset of Fig. 4b), leading to a value of approximately 100, which is larger than the actual experimental system size, confirming that order was spanning the whole experimental system.

Next we use a further privilege of *in silico* investigations —the freedom to change parameter values— and show that asymptotically ordered regimes would probably be observed in slightly different experimental conditions. Experimentally, the vibration amplitude $\Gamma$ was used as control parameter for the onset of collective motion. Decreasing $\Gamma$ to around 2.7 in the experiments, order was observed to increase from near-zero to about $\langle \psi \rangle_t = 0.5$. Unfortunately, due to static friction, the particles stopped moving for $\Gamma$ values below 2.7. To mimic different $\Gamma$-values in the model we multiply both diffusion constants $D_\parallel$ and $D_\perp$ by a coefficient $\gamma^2$, with $\gamma \in [0, 2]$, so that $\gamma = 1$ corresponds to the experiment at $\Gamma = 2.7$. Varying $\gamma$, we find the transition to collective motion to be close to $\gamma = 1$ (Fig. 4b). The transition point is observed to move slightly to the left as the system size is increased. This confirms that vibrated polar disks, in the experimental conditions, are asymptotically disordered, but signals that asymptotically ordered regimes do exist nearby, constituting the first report of long-range orientational order in colliding hard disks without explicit alignment.

Finally, we have performed a systematic exploration of the model varying $\gamma$ and the packing fraction $\phi$ in square domains of linear size $L = 200$ with periodic boundary conditions (Fig. 4c). For $\phi \lesssim 0.6$, varying $\gamma$, we observe the usual phenomenology of models with (effective) polar alignment like the Vicsek model [7, 12, 18, 20, 24]: immediately below the transition, the particles spontaneously segregate in high-density high-order "bands" traveling in a low-density disordered sea (Fig. 4d). Further away from the transition, these nonlinear structures disappear, leaving a statistically-homogeneous Toner-Tu phase with its characteristic giant number fluctuations and long-range correlations [9–11, 25]. However, we detected, for large enough packing fractions, narrow disordered channels (see Fig. 4d, (4)) for *small* noise values (green circles in Fig. 4c). These "inverse bands", not found in dilute or point-like particle models, seem to coexist with the Toner-Tu phase. We believe that the increased frequency of collisions at large packing fractions trigger the emergence of these inhomogeneous structures.

Interestingly, for $\phi \geq 0.6$ we could not observe bands (Fig. 4c). This suggests a possible direct transition from the disordered to the Toner-Tu phase. At this stage, however we cannot conclude, due to numerical limitations, whether this feature remains in the limit of large system sizes and asymptotically large times: the width of the bands increases with increasing $\phi$ (cf. Fig. 4d) so that their disappearance might just be a finite-size effect. However, the longitudinal density profile around $\phi \approx 0.6$ turns out to be rather flat, with an overall rather low order (as low as $\langle \psi \rangle_t \approx 0.2$ for $\phi = 0.6$ and $\gamma = 1.4$). They may thus be of different nature from the Vicsek-like, sharp, well-ordered bands found at low $\phi$, and could cease to exist asymptotically at a packing fraction below the rise of jamming and crystallization effects.

To summarize, we have built a simple yet quantitatively faithful model for the dynamics of the vibrated polar disks studied in [1, 2]. This model constitutes one of the first in which the dynamics of the particle's intrinsic polarity with respect to their velocity is taken into account [56, 57]. An adequate description of the granular system of vibrated discs requires accounting for the polarity as a slow variable compared to the velocity, which can change fast due collisions with the plate or neighboring particles. Our *in silico* study has shown that in the original experiments the most ordered state reached was in fact in the region of the transition to collective motion, slightly on the disordered side. However, asymptotically-ordered regimes do exist nearby. The new features of the phase diagram, i.e. the emergence of "inverse bands" in the low noise regimes of sufficiently dense systems and the possibility of a direct transition from disorder to a collectively-moving Toner-Tu-like phase, deserve further investigations. In particular, this last point, if confirmed in the future, might reopen the debate about the possibility of a continuous transition to collective motion since the structures "responsible" for its discontinuous character —the bands— would then not exist.

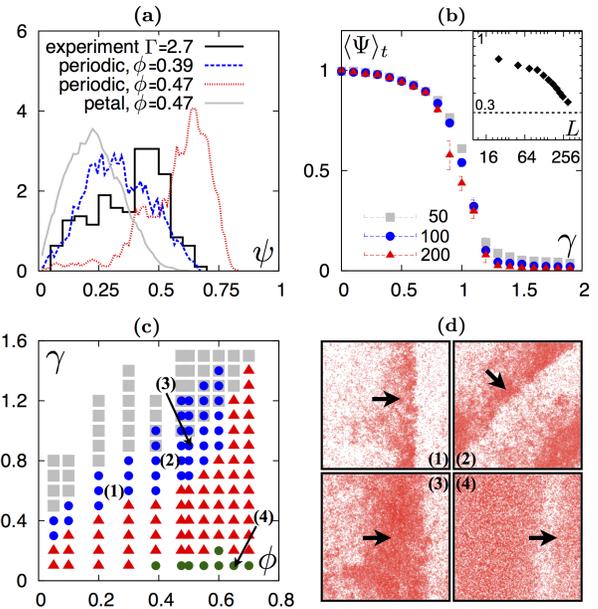

FIG. 4. (color online) **(a)** PDF of the average polarization $\psi$, evaluated within the ROI, for the experimental system, the model in the petal-shaped geometry and in periodic boundaries using two values of packing fractions: $\phi = \{0.39, 0.47\}$. **(b)** Average polarization $\langle \Psi \rangle_t$ as a function of the noise fraction $\gamma^2 = D_\parallel/D_\parallel^{\Gamma=2.7} = D_\perp/D_\perp^{\Gamma=2.7}$, shown for three boundary sizes $L \in \{50, 100, 200\}$ and $\phi = 0.47$. *Inset:* $\langle \Psi \rangle_t$ [log-log] for $\gamma = 1$ and $\phi = 0.47$ as function of system size $L$. **(c)** Sketch of packing fraction($\phi$)-noise($\gamma$) phase diagram: States with $\langle \Psi \rangle_t \leq 0.5$ are indicated by ■, polar homogenous states with $\langle \Psi \rangle_t > 0.5$ by ▲, and states exhibiting heterogenous patterns transversal to the average moving direction ("bands") are depicted by ●. **(d)** Representative snapshots for selected $\phi$-$\gamma$-values indicated by numbers in **(c)**.

E.F. and C.A.W. acknowledge support by the Deutsche Forschungsgemeinschaft in the framework of the SFB 863 "Forces in Biomolecular Systems", and the German Excellence Initiatives via the program "NanoSystems Initiative Munich (NIM)". O.D., J.D., S.L., and H.C. thank the French ANR for financial support (SYSCOMM project DyCoAct).